\newcommand{\derdef}[2]{\ensuremath{\frac{d}{d#2}#1}}

\newcommand{\fmul}[1]{\ensuremath{fmul(#1)}}
\newcommand{\fmulfun}{\ensuremath{fmul}}

\newcommand{\unsat}{\textsc{Unsat}\xspace}
\newcommand{\deltasat}{\textsc{DeltaSat}\xspace}
\newcommand{\maybesat}{\textsc{MaybeSat}\xspace}
\newcommand{\sat}{\textsc{Sat}\xspace}
\newcommand{\unknown}{\textsc{Unknown}\xspace}

\newcommand{\na}{\textsc{n.a.}\xspace}

\newcommand{\pysmt}{\textsc{PySMT}\xspace}
\newcommand{\mathsat}{\textsc{MathSAT}\xspace}
\newcommand{\mathsatnosat}{\textsc{MathSAT-noUniSAT}\xspace}

\newcommand{\isatthree}{\isat}
\newcommand{\metitarski}{\ensuremath{\mbox{\textsc{MetiTarski}}}\xspace}

\newcommand{\dreal}{\textsc{dReal}\xspace}

\newcommand{\hycomp}{\textsc{HyComp}\xspace}

\newcommand{\nuxmv}{\textsc{nuXmv}\xspace}

\newenvironment{small2}{\fontsize{8}{10}\selectfont}{\normalsize}
\newsavebox{\fmbox}
\newenvironment{fmpage}[1]
{\begin{lrbox}{\fmbox}\begin{minipage}{#1}}
{\end{minipage}\end{lrbox}\fbox{\usebox{\fmbox}}}
\newcommand{\defas}{\ensuremath{\stackrel{\text{\tiny def}}{=}}\xspace}
\newcommand{\abst}[1]{\ensuremath{\widehat{#1}}}

\newcommand{\EUF}{\text{UF}\xspace}
\newcommand{\UFLRA}{\text{UFLRA}\xspace}
\newcommand{\NRA}{\text{NRA}\xspace}
\newcommand{\LRA}{\text{LRA}\xspace}
\newcommand{\functionfont}[1]{\textsf{\scalebox{0.9}{#1}}}
\newcommand{\SMTinitialabstraction}{\ensuremath{\functionfont{initial-abstraction}}\xspace}

\newcommand{\SMTUFLRAcheck}{\ensuremath{\functionfont{SMT-\UFLRA-check}}\xspace}
\newcommand{\mktuple}[1]{\ensuremath{\langle #1\rangle}\xspace}

\newcommand{\NTA}{\text{NTA}\xspace}

\newcommand{\SMTNTAcheckCEGAR}{\ensuremath{\functionfont{SMT-\NTA-check-abstract}}\xspace}

\newcommand{\isat}{iSAT3\xspace}

\newcommand{\mksubst}[3]{\mbox{\ensuremath{#1{\{#2 \mapsto #3\}}}}\xspace}

\newcommand{\sref}[1]{\S{}\ref{#1}}
\newcommand{\ignore}[1]{}

\newcommand{\fexp}[1]{\ensuremath{\mathop{f\!exp}(#1)}}
\newcommand{\fsin}[1]{\ensuremath{\mathop{f\!sin}(#1)}}
\newcommand{\ftffun}{\ensuremath{\mathop{f\!t\!f}}}
\newcommand{\ftf}[1]{\ensuremath{\ftffun(#1)}}
\newcommand{\tffun}{\ensuremath{\mathop{t\!f}}}
\newcommand{\tf}[1]{\ensuremath{\tffun(#1)}}

\newcommand{\ub}[1]{\ensuremath{\overline{#1}^{\;u}}}%
\newcommand{\lb}[1]{\ensuremath{\underline{#1}_{\;l}}}%

\newcommand{\tangentLine}[2]{\ensuremath{T_{#1,#2}}\xspace}
\newcommand{\secantLine}[3]{\ensuremath{S_{#1,#2,#3}}\xspace}
\newcommand{\taylorPoly}[2]{\ensuremath{P_{#1,#2}}\xspace}
\newcommand{\taylorRem}[2]{\ensuremath{R_{#1,#2}}\xspace}
\newcommand{\taylorRemUB}[2]{\ub{\ensuremath{R_{#1,#2}}}\xspace}

\documentclass[a4paper]{llncs}
\usepackage{amsmath}
\usepackage{amsfonts}
\usepackage{amssymb}
\usepackage{cite}
\usepackage{xspace}
\usepackage{graphicx}
\usepackage{mathtools}

\usepackage{paralist}

\begin{document}
\sloppypar

\title{%
Satisfiability Modulo Transcendental Functions
via Incremental Linearization\thanks{
This work was funded in part by the H2020-FETOPEN-2016-2017-CSA project SC$^2$ (712689). We thank James Davenport and Erika
Abraham for useful discussions.
}
}
\author{
Alessandro Cimatti\inst{1}\and Alberto Griggio\inst{1} \and Ahmed Irfan\inst{1,2} \and \\ Marco Roveri\inst{1} \and Roberto Sebastiani\inst{2}}
\institute{Fondazione Bruno Kessler, Italy,\\
\email{\texttt{[lastname]}@fbk.eu},\\
\and
DISI, University of Trento, Italy,\\
\email{\texttt{[firstname].[lastname]}@unitn.it}
}

\maketitle

\begin{abstract}
In this paper we present an abstraction-refinement approach to Satisfiability Modulo the
theory of
transcendental functions, such as exponentiation and
trigonometric functions. The transcendental functions are represented
as uninterpreted in the abstract space,
which is described in terms of the combined theory of
linear arithmetic on the rationals with uninterpreted functions,
and are
incrementally axiomatized by means of
upper- and lower-bounding
piecewise-linear functions.
Suitable numerical techniques are used to ensure that the abstractions
of the transcendental functions are sound
even in presence of irrationals.
Our experimental evaluation on benchmarks from verification and
mathematics demonstrates the potential of our approach, showing that
it compares favorably with
delta-satisfiability%
/interval
propagation and methods based on theorem proving.
\end{abstract}
\section{Introduction}
\label{sec:introduction}

Many applications require dealing 
with transcendental functions (e.g.,
exponential, logarithm, sine, cosine). Nevertheless, the problem of
Satisfiability Modulo the theory of 
transcendental functions comes
with many difficulties. First, the problem is in general
undecidable~\cite{transcendental-undecidable}. Second, we may be
forced to deal with irrational numbers - in fact, differently from
polynomial, transcendental functions most often have irrational values for
rational arguments. (See, for example, Hermite's proof that $\exp(x)$ is
irrational for rational non-zero $x$.)

In this paper, we describe a novel approach to Satisfiability
Modulo the quantifier-free theory of 
(nonlinear arithmetic with)
transcendental
functions over the reals - 
SMT(\NTA).
The approach is based on an abstraction-refinement loop, using
SMT(\UFLRA) as abstract space, \UFLRA being the 
combined theory of 
linear arithmetic on the rationals with uninterpreted functions. The Uninterpreted Functions are used
to model nonlinear and transcendental functions. Then, we iteratively
incrementally axiomatize the transcendental functions with a
lemma-on-demand approach.
Specifically, we eliminate spurious interpretations in SMT(\UFLRA) by
tightening the piecewise-linear envelope around the (uninterpreted
counterpart of) the transcendental functions.

A key challenge is to compute provably correct approximations, also in
presence of irrational numbers. 
We use Taylor series to exactly compute
suitable accurate rational coefficients.
We remark that nonlinear polynomials are only used to numerically
compute the coefficients --i.e., no SMT solving in the
  theory of nonlinear arithmetic (SMT(\NRA))  is
  needed--  whereas the refinement is based on the
addition, in the abstract space, of \emph{piecewise-linear} axiom
instantiations, which upper- and lower-bound the candidate solutions, 
ruling out spurious interpretations. 
To compute such piecewise-linear bounding functions,
the concavity of the curve is taken into account.
\ignore{
For example, 
in order to rule out a
spurious interpretation $x=2.0,fexp(x)=3.0$ (where $fexp(x)$ is the
abstraction of the exponential function) we may
exploit the positive concavity of $exp(x)$ and obtain
a linear lower-bound constraint,
e.g. $fexp(x) > \frac{155}{21} + \frac{331}{45}*(x - 2)$.~%
\footnote{%
Notice that $exp(2.0) \approxeq 7.389$,
$\frac{155}{21} \approxeq 7.381\lessapprox exp(2.0)$, and
$\frac{331}{45} \approxeq 7,356\lessapprox \derdef{exp(2.0)}{x}$. 
These values are such that the above linear constraint ``approximates'' the
tangent of $exp(x)$ in $x=2$, 
since it always lower-bounds 
$exp(x)$ and its value and derivative are very near to those of
$exp(x)$ for $x=2.0$.
Recall that $\derdef{exp(x)}{x}=exp(x)$.
}
}
In order to deal with trigonometric functions, we take into account
the periodicity, so that the axiomatization is only done in the
interval between $-\pi$ and $\pi$. Interestingly, not only this is 
helpful for efficiency, but also it is required to ensure correctness.

Another distinguishing feature of our approach is a logical method to
conclude the existence of a solution without explicitly constructing
it.
We use a sufficient criterion that consists in 
checking whether the formula is satisfiable 
under all possible interpretations 
of the uninterpreted functions (representing the transcendental functions)
that are consistent with some rational interval bounds within which
the correct values for the transcendental functions are guaranteed to exist.
We encode the problem as a SMT(\UFLRA) satisfiability check,
such that an unsatisfiable result implies the satisfiability of the original SMT(\NTA) formula.

We implemented the approach on top of the \mathsat SMT
solver~\cite{DBLP:conf/tacas/CimattiGSS13}, using the \pysmt library~\cite{pysmt}.
We experimented with benchmarks from SMT-based
verification queries over nonlinear transition systems, including
Bounded Model Checking of hybrid automata, as well as from several
mathematical properties from the \metitarski~\cite{DBLP:journals/jar/AkbarpourP10} suite and from other competitor solver distributions.
We contrasted our approach with state-of-the-art approaches based on
interval propagation (\isatthree and \dreal), and with the deductive
approach in \metitarski.
The results show that our solver compares favourably with the other solvers,
being able to decide the highest number of benchmarks.

This paper is organized as follows.
In \sref{sec:background} we describe some background.
In \sref{sec:approach} we overview the approach, defining the foundation for safe linear approximations.
In \sref{sec:cegar-transcendental} we describe the specific axiomatization for transcendental functions.
In \sref{sec:related-work} we discuss the related literature,
and in \sref{sec:experiments} we present the experimental evaluation.
In \sref{sec:conclusion} we draw some conclusions and outline directions for future work.
\section{Background}
\label{sec:background}

We assume the standard first-order quantifier-free logical setting and
standard notions of theory, satisfiability, and logical consequence.
As usual in SMT, we denote with \LRA the theory of linear real arithmetic,
with \NRA that of non-linear real arithmetic,
with \EUF the theory of equality (with uninterpreted functions),
and with \UFLRA the combined theory of \EUF and \LRA.
Unless otherwise specified, we use the terms variable and free constant interchangeably.
We denote formulas with $\varphi, \psi$, terms with $t$,
variables with $x, y, a, b$,
functions with $f, \tffun, \ftffun$, each possibly with subscripts.
If $x$ and $y$ are two variables, we denote with $\mksubst{\varphi}{x}{y}$
the formula obtained by replacing all the occurrences of $x$ in $\varphi$ with $y$.
We extend this notation to ordered sequences of variables in the natural way.
If $\mu$ is a model and $x$ is a variable, we write $\mu[x]$ to denote the value of $x$ in $\mu$,
and we extend this notation to terms and formulas in the usual way.
If $\Gamma$ is a set of formulas,
we write $\bigwedge\Gamma$ to denote the formula obtained by taking the conjunction of all its elements.
We write $t_1 < t_2 < t_3$ for $t_1 < t_2 \wedge t_2 < t_3$.

A \emph{transcendental function} is an analytic function that does not
satisfy a polynomial equation
(in contrast to an algebraic function~\cite{townsend2007functions,hazewinkel1993encyclopaedia}).
Within this paper we consider univariate
exponential, logarithmic, and trigonometric functions.
We denote with \NTA the theory of (non-linear) real arithmetic extended with
these transcendental functions.

A \emph{tangent line} to a univariate function $f(x)$ at a point of interest $x=a$ is a
straight line that ``just touches'' the function at the point, and represents the instantaneous rate of change of the function $f$ at that one point.
The tangent line $\tangentLine{f}{a}(x)$ to the function $f$ at point $a$ is the straight line defined as follows:
\begin{equation*}
  \tangentLine{f}{a}(x) \defas f(a) + \derdef{f}{x}(a)*(x - a)
\label{eq:tangent-line}
\end{equation*}
where $\derdef{f}{x}$ is the first-order derivative of $f$ wrt. $x$.

A \emph{secant line} to a univariate function $f(x)$ is a straight line that connects two
points on the function plot.
The secant line $\secantLine{f}{a}{b}(x)$ to a function
$f$ between points $a$ and $b$ is defined as follows:
\begin{equation*}
  \secantLine{f}{a}{b}(x) \defas \frac{f(a) - f(b)}{a - b}*(x - a) + f(a).
\label{eq:secant-line}
\end{equation*}

For a function $f$ that is twice differentiable at point $c$,
the \emph{concavity} of $f$ at $c$ is the sign of its second derivative
evaluated at $c$.
We denote open and closed intervals between two real numbers $l$ and $u$ as $]l, u[$ and $[l, u]$ respectively.
Given a univariate function $f$ over the reals,
the \emph{graph} of $f$ is the set of pairs $\{ \mktuple{x,f(x)}~|~x \in \mathbb{R}\}$.
We might sometimes refer to an element $\mktuple{x,f(x)}$ of the graph as a point.

\paragraph{Taylor Series and Taylor's Theorem.}
Given a function $f(x)$ that has $n+1$ continuous derivatives at
$x=a$, the \emph{Taylor series} of degree $n$ centered around $a$ is the polynomial:
\begin{equation*}
\label{eq:taylor1}
\taylorPoly{n}{f(a)}(x) \defas \sum_{i=0}^{n} \frac{f^{(i)}(a)}{i!} * (x - a)^i
\end{equation*}
where $f^{(i)}(a)$ is the evaluation of $i$-th derivative of $f(x)$ at
point $x=a$. The Taylor series centered around $0$ is also called
\emph{Maclaurin series}.

According to \emph{Taylor's theorem}, any continuous function $f(x)$ that
is $n+1$ differentiable can be written as the sum of the Taylor series and
the remainder term:
\begin{equation*}
\label{eq:taylor2}
f(x) = \taylorPoly{n}{f(a)}(x) + \taylorRem{n+1}{f(a)}(x)
\end{equation*}
where $\taylorRem{n+1}{f(a)}(x)$ is basically the Lagrange form of the
remainder, and for some point $b$ between $x$ and $a$ it is given by:
\begin{equation*}
\label{eq:taylor3}
  \taylorRem{n+1}{f(a)}(x) \defas \frac{f^{(n+1)}(b)}{(n+1)!}*(x-a)^{n+1}.
\end{equation*}
The value of the point $b$ is not known, but the upper bound on the
size of the remainder $\taylorRemUB{n+1}{f(a)}(x)$ at a point $x$ can be
estimated by:
\begin{equation*}
\taylorRemUB{n+1}{f(a)}(x) \defas \max_{c \in [\min(a, x), \max(a, x)]}(|f^{(n+1)}(c)|) * \frac{|(x-a)^{n+1}|}{(n+1)!}.
\end{equation*}
This allows to obtain two polynomials
that are above and below the function at a given point $x$, by considering
$\taylorPoly{n}{f(a)}(x) + \taylorRemUB{n+1}{f(a)}(x)$ and
$\taylorPoly{n}{f(a)}(x) - \taylorRemUB{n+1}{f(a)}(x)$ respectively.

\section{Overview of the approach}
\label{sec:approach}

\newcounter{pseudocodecounter}
\newcommand{\pcl}{%
  \refstepcounter{pseudocodecounter}
  {\scriptsize \arabic{pseudocodecounter}.}\hspace{1em}%
}
\newcommand{\pcll}[1]{\pcl\label{#1}} %
\newcommand{\pcc}[1]{%
  {\it ~\# #1}%
}
\newcommand{\pcs}{%
  {\phantom {\scriptsize \arabic{pseudocodecounter}.}\hspace{1em}}
}

\newcommand{\codeIte}[3]{\ensuremath{(#1)~?~(#2)~:~(#3)}}
\newcommand{\codeProcName}[1]{\ensuremath{\text{\textsf{#1}}}\xspace}
\newcommand{\checkRefine}{\codeProcName{check-refine}}
\newcommand{\checkRefinePoly}{\codeProcName{check-refine-NRA}}
\newcommand{\refinePoint}{\codeProcName{get-lemmas-point}}
\newcommand{\polyApprox}{\codeProcName{poly-approx}}
\newcommand{\getConcavity}{\codeProcName{get-concavity}}
\newcommand{\getTangentBounds}{\codeProcName{get-tangent-bounds}}
\newcommand{\checkModel}{\codeProcName{check-model}}
\newcommand{\getBounds}{\codeProcName{get-bounds}}
\newcommand{\getPreviousRefinementPoints}{\codeProcName{get-previous-secant-points}}
\newcommand{\storeRefinementPoint}{\codeProcName{store-secant-point}}
\newcommand{\budgetExhausted}{\codeProcName{budget-exhausted}}
\newcommand{\initialPrecision}{\codeProcName{initial-precision}}
\newcommand{\maybeIncreasePrecision}{\codeProcName{maybe-increase-precision}}
\newcommand{\refineExtra}{\codeProcName{refine-extra}}

\begin{figure}[t]
  \centering
    \begin{fmpage}{1.0\linewidth}
  \begin{small2}
    \setcounter{pseudocodecounter}{0}
      \begin{tabbing}
        {\bf bool} \SMTNTAcheckCEGAR($\varphi$):\\
        \pcs xxx \= xxx \= xxx \= xxx \= xxx \= xxx \= \kill
        \pcl $\abst{\varphi} = \SMTinitialabstraction(\varphi)$\\
        \pcl $\Gamma = \emptyset$\\
        \pcl precision := \initialPrecision()\\
        \pcl {\bf while true}:\\
        \pcl \> {\bf if} \budgetExhausted(): \\
        \pcl \> \> {\bf abort}\\
        \pcll{code:UF-LRA-check} \> \mktuple{res, \abst{\mu}} = \SMTUFLRAcheck($\abst{\varphi} \land \bigwedge \Gamma$) \\
        \pcl \> {\bf if not} res: \\
        \pcl \> \> {\bf return false} \\
        \pcl \> \mktuple{sat, \Gamma'} := \checkRefine($\varphi$, $\abst{\mu}$, precision)\\
        \pcl \> {\bf if} sat: \\
        \pcl \> \> {\bf return true}\\
        \pcl \> {\bf else}: \\
        \pcl \> \> precision := \maybeIncreasePrecision()\\
        \pcl \> \> $\Gamma''$ := \refineExtra($\varphi$, $\abst{\mu}$)\\
        \pcl \>\> $\Gamma$ := $\Gamma \cup \Gamma' \cup \Gamma''$
      \end{tabbing}
  \end{small2}
    \end{fmpage}
  \caption{Solving SMT(\NTA) via abstraction to SMT(\UFLRA).
    \label{fig:smt-nta-pseudocode}}
\end{figure}

Our procedure, which extends to SMT(\NTA) and pushes further the approach
 presented in~\cite{tacas17} for SMT(\NRA),
works  by overapproximating the input formula with a
formula over the combined theory of linear arithmetic and
uninterpreted functions.
The main algorithm is shown in Fig.~\ref{fig:smt-nta-pseudocode}.
The solving procedure follows a classic abstraction-refinement loop,
in which at each iteration, the current safe approximation $\abst{\varphi}$ of the
input SMT(\NTA) formula $\varphi$ is refined by adding new
constraints $\Gamma$ that rule out one (or possibly more) spurious solutions,
until one of the following conditions occurs:
\begin{inparaenum}[(i)]
\item the resource budget (e.g. time, memory, number of iterations) is
  exhausted; or
\item $\abst{\varphi} \land \bigwedge \Gamma$ becomes unsatisfiable in SMT(\UFLRA);
  or
\item the SMT(\UFLRA) satisfiability result for $\abst{\varphi} \land
  \bigwedge \Gamma$ can be lifted to a satisfiability result for the
  original formula $\varphi$.
\end{inparaenum}
An initial current precision is set (calling the function \initialPrecision),
and this value is possibly increased at each iteration (calling \maybeIncreasePrecision)
according to the result of \checkRefine and some heuristic.

In Fig.~\ref{fig:smt-nta-pseudocode} we distinguish between two different refinement procedures:
\begin{inparaenum}[1)]
\item \checkRefine, which is described below;
\item \refineExtra, which is  described in \sref{sec:cegar-transcendental}, where we provide further details on the treatment
  of each specific transcendental function that we currently support.
\end{inparaenum}
\subsubsection{Initial Abstraction.}
The function \SMTinitialabstraction takes in input an SMT(\NTA) formula
$\varphi$ and returns a SMT(\UFLRA) safe approximation
$\abst{\varphi}$ of it.
First, we flatten each transcendental function application $\tf{t}$
in $\varphi$ in which $t$ is not a variable
by replacing $t$ with a fresh variable $y$, and by
conjoining $y = t$ to $\varphi$.
Then, we replace each transcendental function $\tf{x}$
in $\varphi$ with a corresponding uninterpreted function $\ftf{x}$,
producing thus an SMT(\UFLRA) formula $\abst{\varphi}$.
Finally, we add to $\abst{\varphi}$
some simple initial axioms for the different transcendental
functions, expressing general, simple mathematical properties about them.
We shall describe such axioms in \sref{sec:cegar-transcendental}.

If $\varphi$ contains also non-linear polynomials,
we handle them as described in \cite{tacas17}:
we replace each non-linear product $t_1*t_2$ with an uninterpreted function application $\fmul{t_1, t_2}$,
and add to the input formula some initial axioms
expressing general, simple mathematical properties of
multiplications. (We refer the reader to  \cite{tacas17} for details.)

\subsubsection{Spuriousness check and abstraction refinement.}
\begin{figure}[t]
  \centering
    \begin{fmpage}{1.0\linewidth}
  \begin{small2}
    \setcounter{pseudocodecounter}{0}
      \begin{tabbing}
        \mktuple{{\bf bool}, \text{axiom-set}} \checkRefine($\varphi$, $\abst{\mu}$, precision):\\
        \pcs xxx \= xxx \= xxx \= xxx \= xxx \= xxx \= \kill
        \pcll{check-refine-tacas17} $\Gamma$ := \checkRefinePoly($\varphi$, $\abst{\mu}$) \pcc{NRA refinement of \cite{tacas17}}\\%
        \pcl $\epsilon$ := $10^{-\text{precision}}$\\
        \pcll{refine-begin-for} {\bf for all} $\tf{x} \in \varphi$:\\
        \pcl \> $c$ := $\abst{\mu}[x]$\\
        \pcl \> $\mktuple{P_l(x), P_u(x)}$ := \polyApprox($\tf{x}$, $c$, $\epsilon$)\\
        \pcl \> {\bf if} $\abst{\mu}[\ftf{x}] \le P_l(c)$ {\bf or} $\abst{\mu}[\ftf{x}] \ge P_u(c)$:\\
        \pcll{call-refine-point}\label{refine-end-for} \> \> $\Gamma$ := $\Gamma \cup$ \refinePoint($\tf{x}$, $\abst{\mu}$, $P_l(x)$, $P_u(x)$)\\
        \pcl {\bf if} $\Gamma = \emptyset$:\\
        \pcll{call-check-model} \> {\bf if} \checkModel($\varphi$, $\abst{\mu}$):\\
        \pcll{refine-return-lemmas} \> \> {\bf return} \mktuple{{\bf true}, \emptyset}\\
        \pcl \> {\bf else}:\\
        \pcll{repeat-check-refine} \> \> {\bf return} \checkRefine($\varphi$, $\abst{\mu}$, precision+1)\\
        \pcl {\bf else}:\\
        \pcl \> {\bf return} \mktuple{{\bf false}, \Gamma}
      \end{tabbing}
  \end{small2}
    \end{fmpage}
  \caption{The main refinement procedure.
    \label{fig:smt-nta-refine}}
\end{figure}
The core of our procedure is the $\checkRefine$ function,
shown in Fig.~\ref{fig:smt-nta-refine}.

First, %
if the formula contains also some non-linear polynomials,
\checkRefine performs the refinement of non-linear multiplications
as described in~\cite{tacas17}.
In Fig.~\ref{fig:smt-nta-refine}, this is represented by the call to
the function \checkRefinePoly at line~\ref{check-refine-tacas17},
which may return some axioms to further constrain $\fmulfun$ terms.
If no non-linear polynomials occur in $\varphi$, then $\Gamma$ is
initialized as the empty set.

Then, the function iterates over all the transcendental function
applications $\tf{x}$ in $\varphi$ (lines
\ref{refine-begin-for}--\ref{refine-end-for}), and checks whether the
SMT(\UFLRA)-model $\abst{\mu}$ is consistent with their semantics.

Intuitively, in principle, this amounts to check that $\tf{\abst{\mu}[x]}$ is equal to $\abst{\mu}[\ftf{x}]$.
In practice, however, the check cannot be exact,
since transcendental functions at rational points typically have irrational values (see e.g. \cite{nieven}),
which cannot be represented in SMT(\UFLRA).
Therefore, for each $\tf{x}$ in $\varphi$, we instead compute two polynomials,
$P_l(x)$ and $P_u(x)$, with the property that $\tf{\abst{\mu}[x]}$
belongs to the open interval $]P_l(\abst{\mu}[x]), P_u(\abst{\mu}[x])[$.
The polynomials are computed using Taylor series,
according to the given current precision, by the function \polyApprox,
which shall be described in \sref{sec:cegar-transcendental}.

If the model value $\abst{\mu}[\ftf{x}]$ for $\tf{x}$ is outside the above interval,
then the function \refinePoint is used to generate some linear lemmas
that will remove the spurious point $\mktuple{\abst{\mu}[x], \abst{\mu}[\ftf{x}]}$
from the graph of the current abstraction of $\tf{x}$ (line \ref{call-refine-point}).

If at least one point was refined in the loop of lines \ref{refine-begin-for}--\ref{refine-end-for},
the current set of lemmas $\Gamma$ is returned (line \ref{refine-return-lemmas}).
If instead none of the points was determined to be spurious,
the function \checkModel is called (line \ref{call-check-model}).
This function tries to determine whether the abstract model $\abst{\mu}$
does indeed imply the existence of a model for the original formula $\varphi$
(more details are given below).
If the check fails, we repeat the \checkRefine call with an increased precision (line \ref{repeat-check-refine}).

\subsubsection{Refining a spurious point with secant and tangent lines.}

\begin{figure}[t]
  \centering
    \begin{fmpage}{1.0\linewidth}
  \begin{small2}
    \setcounter{pseudocodecounter}{0}
      \begin{tabbing}
        axiom-set \refinePoint($\tf{x}$, $\abst{\mu}$, $P_l(x)$, $P_u(x)$):\\
        \pcs xxx \= xxx \= xxx \= xxx \= xxx \= xxx \= \kill
        \pcl $c$ := $\abst{\mu}[x]$ \\
        \pcl $v$ := $\abst{\mu}[\ftf{x}]$ \\
        \pcl conc := \getConcavity($\tf{x}$, $c$)\\
        \pcll{tangent-lemma-begin} {\bf if} ($v \le P_l(c)$ {\bf and} $\text{conc} \ge 0$) {\bf or}
                      ($v \ge P_u(c)$ {\bf and} $\text{conc} \le 0$):\\
        \pcs \pcc{tangent refinement}\\
        \pcl \> $P$ := $\codeIte{v \le P_l(c)}{P_l}{P_u}$\\
        \pcl \> $T(x)$ := $P(c) + \derdef{P}{x}(c) \cdot (x - c)$ \pcc{tangent of $P$ at $c$}\\
        \pcll{get-tangent-bounds} \> \mktuple{l, u} := \getTangentBounds($\tf{x}$, $c$, $\derdef{P}{x}(c)$) \\
        \pcl \> $\psi$ := $\codeIte{\text{conc} < 0}{\ftf{x} \le T(x)}{\ftf{x} \ge T(x)}$ \\
        \pcll{tangent-lemma-end} \> {\bf return} \{$((x \ge l) \land (x \le u)) \rightarrow \psi$\}\\
        \pcll{secant-lemma-begin} {\bf else}: \pcc{$(v \le P_l(c) \land \text{conc} < 0) \lor (v \ge P_u(c) \land \text{conc} > 0)$}\\
        \pcs \pcc{secant refinement}\\
        \pcl \> prev := \getPreviousRefinementPoints($\tf{x}$)\\
        \pcl \> $l$ := $\text{max} \{ p \in \text{prev} ~|~ p < c \}$ \\
        \pcl \> $u$ := $\text{min} \{ p \in \text{prev} ~|~ p > c \}$ \\
        \pcl \> $P$ := $\codeIte{v \le P_l(c)}{P_l}{P_u}$\\
        \pcl \> $S_l(x)$ := $\dfrac{P(l)-P(c)}{l-c}\cdot (x - l) + P(l)$ \pcc{secant of $P$ between $l$ and $c$}\\
        \pcl \> $S_u(x)$ := $\dfrac{P(u)-P(c)}{u-c}\cdot (x - u) + P(u)$\\
        \pcl \> $\psi_l$ := $\codeIte{\text{conc} < 0}{\ftf{x} \ge S_l(x)}{\ftf{x} \le S_l(x)}$ \\
        \pcl \> $\psi_u$ := $\codeIte{\text{conc} < 0}{\ftf{x} \ge S_u(x)}{\ftf{x} \le S_u(x)}$ \\
        \pcl \> $\phi_l$ := $(x \ge l) \land (x \le c)$ \\
        \pcl \> $\phi_u$ := $(x \ge c) \land (x \le u)$ \\
        \pcl \> \storeRefinementPoint($\tf{x}$, $c$)\\
        \pcll{secant-lemma-end} \> {\bf return}
        \{$(\phi_l \rightarrow \psi_l)$, $(\phi_u \rightarrow \psi_u)$\}
      \end{tabbing}
  \end{small2}
    \end{fmpage}
  \caption{Piecewise-linear refinement for the transcendental function $\tf{x}$ at point $c$.
    \label{fig:smt-nta-refine-point}}
\end{figure}

\begin{figure}[t]
  \centering
  \includegraphics[width=0.6\textwidth,trim={0 1.5mm 0 1.5mm},clip]{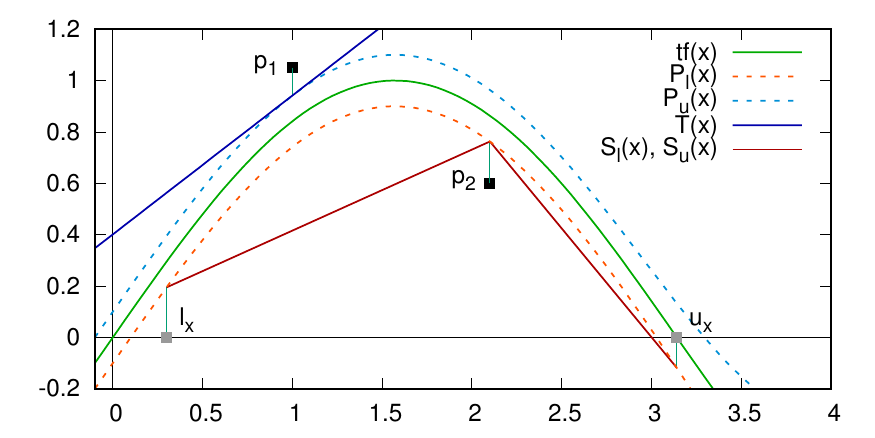}
  \caption{Piecewise-linear refinement illustration.
    \label{fig:tangent-secant-approx-example}}
\end{figure}

Given a transcendental function application $\tf{x}$,
the \refinePoint function generates a set of lemmas for refining the interpretation of $\ftf{x}$
by constructing a piecewise-linear approximation of $\tf{x}$ around the point $\abst{\mu}[x]$,
using one of the polynomials $P_l(x)$ and $P_u(x)$ computed in \checkRefine.
The kind of lemmas generated, and which of the two polynomials is used,
depend on
\begin{inparaenum}[(i)]
\item the position of the spurious value $\abst{\mu}[\ftf{x}]$ relative to the correct value $\tf{\abst{\mu}[x]}$, and
\item the concavity of $\tffun$ around the point $\abst{\mu}[x]$.
\end{inparaenum}
If the concavity is positive (resp. negative) or equal to zero, and the point lies below (resp. above) the function,
then the linear approximation is given by a tangent
to the lower (resp. upper) bound polynomial $P_l$ (resp. $P_u$)
at $\abst{\mu}[x]$ (lines \ref{tangent-lemma-begin}--\ref{tangent-lemma-end} of Fig.~\ref{fig:smt-nta-refine-point});
otherwise, i.e. the concavity is negative (resp. positive)
and the point is below (resp. above) the function,
the linear approximation is given by a pair of secants
to the lower (resp. upper) bound polynomial $P_l$ (resp. $P_u$)
around $\abst{\mu}[x]$ (lines \ref{secant-lemma-begin}--\ref{secant-lemma-end} of Fig.~\ref{fig:smt-nta-refine-point}).
The two situations are illustrated in Fig.~\ref{fig:tangent-secant-approx-example}.

In the case of tangent refinement, the function \getTangentBounds (line \ref{get-tangent-bounds})
returns an interval $[l, u]$ such that the tangent line is guaranteed not to cross the transcendental function \tffun.
In practice, this interval can be (under)approximated quickly by exploiting
known properties of the specific function $\tffun$ under consideration.
For example, for the exponential function \getTangentBounds always returns $[-\infty, +\infty]$;
for other functions, the computation can be based e.g. on an analysis of the
(known, precomputed) inflection points of $\tffun$
around the point of interest $\abst{\mu}[x]$
and the slope $\derdef{P}{x}(c)$ of the tangent line.

In the case of secant refinement,
a second value, different from $\abst{\mu}[x]$,
is required to draw a secant line.
The function \getPreviousRefinementPoints returns the set of all the points
at which a secant refinement was performed in the past for $\tf{x}$.
From this set, we take the two points closest to $\abst{\mu}[x]$, such that
$l < \abst{\mu}[x] < u$ and that $l,u$ do not cross any inflection point,
\footnote{For simplicity, we assume that this is always possible.
If needed, this can be implemented e.g. by generating
the two points at random while ensuring that $l < \abst{\mu}[x] < u$ and that $l,u$ do not cross any inflection point.}
and use those points to generate two secant lines and their validity intervals.
Before returning the set of the two corresponding lemmas, we also store
the new secant refinement point $\abst{\mu}[x]$ by calling \storeRefinementPoint.

\subsubsection{Detecting satisfiable formulas.}

The function \checkModel tries to determine whether the \UFLRA-model
$\abst{\mu}$ for $\abst{\varphi} \land \bigwedge \Gamma$
implies the satisfiability of the original formula $\varphi$.
If, for all $\tf{x}$ in $\varphi$,
\tffun{} has a rational value at the rational point $\abst{\mu}[x]$,%
\footnote{Although, as mentioned above, this is not the case in general (see e.g. \cite{nieven}),
it is true for some special values, e.g. $\exp(0) = 1$, $\sin(0) = 0$.}
and $\abst{\mu}[\ftf{x}]$ is equal to $\tf{\abst{\mu}[x]}$,
then $\abst{\mu}$ can be directly lifted to a model $\mu$ for $\varphi$.

In the general case, we exploit this simple observation:
we can still conclude that $\varphi$ is satisfiable
if we are able to show that $\abst{\varphi}$ is satisfiable
\emph{under all possible interpretations of \ftffun}
that are guaranteed to include also \tffun.

Using the model $\abst{\mu}$, we compute safe lower and upper bounds
$\lb{\tf{\abst{\mu}[x]}}$ and $\ub{\tf{\abst{\mu}[x]}}$
for the function $\tffun$ at point $\abst{\mu}[x]$
with the \polyApprox function (see above).
Let $FTF$ be the set of all $\ftf{x}$ terms occurring in
$\abst{\varphi}$.
Let $V$ be the set of variables $x$ for $\ftf{x} \in FTF$,
and $F$ be the set of all the function symbols in $FTF$.
Intuitively, if we can prove the validity of the following formula:

\vskip -\baselineskip%
\begin{small}
  \begin{equation*}
\forall \ftffun \in F.
  \left(\bigwedge_{{\ftf{x} \in FTF}}
    \lb{\tf{{\abst{\mu}[x]}}} \le \ftf{{\abst{\mu}[x]}} \le \ub{\tf{{\abst{\mu}[x]}}}
  \right) \rightarrow \mksubst{\abst{\varphi}}{V}{\abst{\mu}[V]} \label{eq:allinterpretations}
\end{equation*}
\end{small}
\vskip -\baselineskip%

\noindent then the original formula $\varphi$ is satisfiable.

In order to be able to use a quantifier-free SMT(\UFLRA)-solver,
we reduce the problem to the validity check of a pure \UFLRA formula.
Let $CT$ be the set of all terms $\ftf{{\abst{\mu}[x]}}$ occurring in $\mksubst{\abst{\varphi}}{V}{\abst{\mu}[V]}$.
We replace each occurrence of $\ftf{{\abst{\mu}[x]}}$ in $CT$
with a corresponding fresh variable $y_{\ftf{{\abst{\mu}[x]}}}$ from a set $Y$.
We then check the validity of the formula:

\vskip -\baselineskip%
\begin{small}
\begin{equation*}
  \varphi_{\abst{\mu}}^{\text{sat}} \defas
  \forall Y.
  \mksubst{\displaystyle\left(\!\!\left(\bigwedge_{{\ftf{x} \in FTF}}\!
    \lb{\tf{{\abst{\mu}[x]}}} \le \ftf{{\abst{\mu}[x]}} \le \ub{\tf{{\abst{\mu}[x]}}}
  \!\right)\! \rightarrow \mksubst{\abst{\varphi}}{V}{\abst{\mu}[V]}\!\right)}{CT}{Y}.
\end{equation*}
\end{small}
\vskip -\baselineskip%

If $\neg \varphi_{\abst{\mu}}^{\text{sat}}$ is unsatisfiable,
we conclude that $\varphi$ is satisfiable.
Clearly, this can be checked with a quantifier-free
SMT(\UFLRA)-solver, %
 since $\neg \forall x.\phi$ is equivalent to $\exists x.\neg \phi$,
and $x$ can then be removed by Skolemization.
\section{Abstraction Refinement for Transcendental Functions}
\label{sec:cegar-transcendental}

In this section, we describe the implementation of the
\polyApprox and \refineExtra
for the transcendental functions that we currently support,
namely $\exp$ and $\sin$.%
\footnote{We remark that our tool (see \S\ref{sec:experiments}) can handle also
$\log$, $\cos$, $\tan$, $\arcsin$, $\arccos$, $\arctan$ by means of rewriting.
We leave as future work the possibility of handling such functions natively.
}

The $\polyApprox(\tf{x}, c, \epsilon)$ function
uses the Maclaurin series of the corresponding transcendental function
and Taylor's theorem
to find the lower and upper polynomials.
Essentially, this is done by expanding the series
(and the remainder approximation)
up to a certain $n$,
until the desired precision $\epsilon$
(i.e. the difference between the upper and lower polynomials evaluated at $c$)
is achieved.
Notice that, since
we can precisely evaluate the derivative of any order at $0$
for both $\exp$ and $\sin$,%
\footnote{Because
  \begin{inparaenum}[(i)]
    \item $\exp(0)=1$, $\sin(0)=0$, $\cos(0)=1$,
    \item $\exp^{(i)}(x) = \exp(x)$ for all $i$, and
    \item $|\sin^{(i)}(x)|$ is $|\cos(x)|$ if $i$ is odd and $|\sin(x)|$ otherwise.
  \end{inparaenum}}
the computation of both the Maclaurin series and the remainder polynomial is always exact.
\subsubsection{Exponential Function}

\paragraph{Piecewise-Linear Refinement.}
The polynomial $\taylorPoly{n}{\exp(0)}(x)$ given by the Maclaurin
series behaves differently depending on the sign of $x$. For that
reason, \polyApprox distinguishes three cases for finding the polynomials $P_l(x)$ and $P_u(x)$:
\begin{description}
\item[Case $x = 0$:] since $\exp(0)=1$, we have $P_l(0) = P_u(0) = 1$;
\item[Case $x < 0$:] we have that $\taylorPoly{n}{\exp(0)}(x) < \exp(x)$
  if $n$ is odd, and $\taylorPoly{n}{\exp(0)}(x) > \exp(x)$ if $n$ is even
  (where $ \taylorPoly{n}{\exp(0)}(x) = \sum_{i=0}^{n} \frac{x^i}{i!}$);
  we therefore set $P_l(x)=\taylorPoly{n}{\exp(0)}(x)$ and $P_u(x) = \taylorPoly{n+1}{\exp(0)}(x)$ for a suitable $n$ so that the required precision $\epsilon$ is met;
\item[Case $x>0$:] we have that $\taylorPoly{n}{\exp(0)}(x) < \exp(x)$
  and $\taylorPoly{n}{\exp(0)}(x)*(1 - \frac{x^{n+1}}{(n+1)!})^{-1} > \exp(x)$ when $(1 - \frac{x^{n+1}}{(n+1)!}) > 0$,
  therefore we set $P_l(x) = \taylorPoly{n}{\exp(0)}(x)$ and $P_u(x) = \taylorPoly{n}{\exp(0)}(x)*(1 - \frac{x^{n+1}}{(n+1)!})^{-1}$ for a suitable $n$.
\end{description}

Since the concavity of $\exp$ is always positive,
the tangent refinement will always give lower bounds for $\exp(x)$,
and the secant refinement will give upper bounds.
Moreover, as $\exp$ has no inflection points, \getTangentBounds always returns $[-\infty,+\infty]$.

\paragraph{Extra Refinement.}
The exponential function is monotonically increasing with a non-linear order.
We check this property between two $\fexp{x}$ and $\fexp{y}$ terms in $\abst{\varphi}$:
if $\abst{\mu}[x] < \abst{\mu}[y]$,
but $\abst{\mu}[\fexp{x}] \not< \abst{\mu}[\fexp{y}]$,
then we add the following extra refinement lemma:
\begin{displaymath}
x < y \leftrightarrow \fexp{x} < \fexp{y}
\end{displaymath}

\paragraph{Initial Axioms.}
We add the following initial axioms to $\abst{\varphi}$.
\begin{align*}
\textit{Lower Bound: }         & \fexp{x} > 0\\
\textit{Zero: }                & (x=0 \leftrightarrow \fexp{x} = 1) \wedge%
                                 (x < 0 \leftrightarrow \fexp{x} < 1) \wedge\\%
                               & (x > 0 \leftrightarrow \fexp{x} > 1)\\
\textit{Zero Tangent Line: }& x=0 \vee \fexp{x} > x + 1
\end{align*}

\subsubsection{Sin Function}

\paragraph{Piecewise-Linear Refinement.}

The correctness of our refinement procedure relies crucially on being able
to compute the concavity of the transcendental function $\tffun$ at a given point $c$.
This is needed in order to know whether a computed tangent or secant line
constitutes a valid upper or lower bound for $\tffun$ around $c$ (see Fig.~\ref{fig:smt-nta-refine-point}).
In the case of the $\sin$ function,
computing the concavity at an arbitrary point $c$ is problematic,
since this essentially amounts to computing the remainder
of $c$ and $\pi$, which,
being $\pi$ a transcendental number, cannot be exactly computed.

In order to solve this problem,
we exploit another property of $\sin$, namely its periodicity (with period $2\pi$).
More precisely, we split the reasoning about $\sin$
depending on two kinds of periods: base period and extended period.
A period is a \emph{base period} for the $\sin$ function if it is from
$-\pi$ to $\pi$, otherwise it is an \emph{extended period}.
In order to reason about periods,
we first introduce a symbolic variable $\abst{\pi}$,
and add the constraint $l_\pi < \abst{\pi} < u_\pi$ to $\abst{\varphi}$,
where $l_\pi$ and $u_\pi$ are valid rational lower and upper bounds for the actual value of $\pi$
(in our current implementation, we have $l_\pi = \frac{333}{106}$ and $u_\pi = \frac{355}{113}$).
Then, we introduce for each \fsin{x} term an ``artificial'' $\sin$
function application \fsin{y_x} (where $y_x$ is a fresh variable),
 whose domain is the base period. This is done by adding the following
constraints:
\begin{displaymath}
(-\abst{\pi} \leq y_x \leq \abst{\pi}) \wedge%
((-\abst{\pi} \leq x \leq \abst{\pi}) \rightarrow y_x = x) \wedge%
\fsin{x} = \fsin{y_x}.
\end{displaymath}
We call these fresh variables \emph{base variables}.
Notice that the second and the third constraint are saying
that \fsin{x} is the same as \fsin{y_y} in the base period.

Let $F\!sin_{base}$ be the set of \fsin{y_x} terms that have base variables
as arguments, $F\!sin$ be the set of all \fsin{x} terms,
and
$F\!sin_{ext} \defas F\!sin - F\!sin_{base}$.
The tangent and secant refinement is performed for the terms in $F\!sin_{base}$,
while we add a \emph{linear shift} lemma (described below)
as refinement for the terms in $F\!sin_{ext}$.
Using this transformation, we can easily compute the concavity of $\sin$ at
$\abst{\mu}[y_x]$ by just looking at the sign of $\abst{\mu}[y_x]$,
\emph{provided that $-l_\pi \leq \abst{\mu}[y_x] \leq l_\pi$},
where $l_\pi$ is the current lower bound for $\abst{\pi}$.%
\footnote{In the interval $[-\pi, \pi]$, the concavity of $\sin(c)$ is the opposite of the sign of $c$.}
In the case in which $-u_\pi < \abst{\mu}[y_x] < -l_\pi$ or $l_\pi < \abst{\mu}[y_x] < u_\pi$,
we do not perform the tangent/secant refinement,
but instead we refine the precision of $\abst{\pi}$.
For each $\fsin{y_x} \in FSin_{base}$, \polyApprox tries to find the
lower and upper polynomial using Taylor's theorem,
which ensures that:
\begin{displaymath}
  \taylorPoly{n}{\sin(0)}(y_x) -
  \taylorRemUB{n+1}{\sin(0)}(y_x) \leq \sin(y_x)
  \leq \taylorPoly{n}{\sin(0)}(y_x) +
  \taylorRemUB{n+1}{\sin(0)}(y_x)
\end{displaymath}
where
$\taylorPoly{n}{\sin(0)}(y_x) = \sum_{k=0}^n \frac{(-1)^k * y_x^{2k+1}}{(2k +1)!}$
and
$\taylorRemUB{n+1}{\sin(0)}(y_x) = \frac{y_x^{2(n+1)}}{(2(n+1))!}$.
Therefore, we can set $P_l(x) = \taylorPoly{n}{\sin(0)}(x) - \taylorRemUB{n+1}{\sin(0)}(x)$
and $P_u(x) = \taylorPoly{n}{\sin(0)}(x) + \taylorRemUB{n+1}{\sin(0)}(x)$.

\paragraph{Extra Refinement.}
For each $\fsin{x} \in F\!sin_{ext}$
with the corresponding base variable $y_x$,
we check whether the value $\abst{\mu}[x]$ after shifting to the base period
is equal to the value of $\abst{\mu}[y_x]$.
We calculate the shift $s$ of $x$ as the rounding towards zero
of $(\abst{\mu}[x]+\abst{\mu}[\abst{\pi}])/(2\cdot\abst{\mu}[\abst{\pi}])$, and
we then compare $\abst{\mu}[y_x]$ with $\abst{\mu}[x] - 2s\cdot\abst{\mu}[\abst{\pi}]$.
If the values are different, we add the following \emph{shift lemma}
for relating $x$ with $y_x$ in the extended period $s$:
\begin{displaymath}
  (\abst{\pi} * (2s - 1) \leq x \leq \abst{\pi} * (2s + 1)) \rightarrow y_x = x - 2s*\abst{\pi}.
\end{displaymath}
In this way, we do not need the tangent and secant refinement
for the extended period and we can reuse the refinements done in the
base period.
Note that even if the calculated shift value is wrong
(due to the imprecision of $\abst{\mu}[\abst{\pi}]$ with respect to the real value $\pi$),
we may generate something useless but never wrong.

We also check the monotonicity property of $\sin$, which can be
described for the base period as: (i) the $\sin$ is monotonically
increasing in the interval $-\frac{\pi}{2}$ to $\frac{\pi}{2}$; (ii)
the $\sin$ is monotonically decreasing in the intervals $-\pi$ to
$-\frac{\pi}{2}$ and $\frac{\pi}{2}$ to $\pi$.
We add one of the constraints below if it is in conflict according to
the current abstract model for some $\fsin{y_{x_1}},\fsin{y_{x_2}} \in
F\!sin_{base}$.
\begin{align*}
(-\frac{\abst{\pi}}{2} \leq y_{x_1} < y_{x_2} \leq \frac{\abst{\pi}}{2}) \rightarrow fsin(y_{x_1}) < fsin(y_{x_2})\\
(-\abst{\pi} \leq y_{x_1} < y_{x_2} \leq -\frac{\abst{\pi}}{2}) \rightarrow fsin(y_{x_1}) > fsin(y_{x_2})\\
(\frac{\abst{\pi}}{2} \leq y_{x_1} < y_{x_2} \leq \abst{\pi}) \rightarrow fsin(y_{x_1}) > fsin(y_{x_2})
\end{align*}

\paragraph{Initial Axioms.}
For each $\fsin{z} \in F\!sin$, we add the generic lower and upper
bounds: $-1 \leq \fsin{z} \leq 1$.
For each $\fsin{y_x} \in F\!sin_{base}$, we add the following axioms.
\begin{align*}
\textit{Symmetry: }        & \fsin{y_x} = - \fsin{-y_x}\\
\textit{Phase: }           & (0 < y_x < \abst{\pi} \leftrightarrow \fsin{y_x} > 0) \wedge%
                             (-\abst{\pi} < y_x < 0 \leftrightarrow \fsin{y_x} < 0) \\
\textit{Zero Tangent: }    & (y_x > 0 \rightarrow \fsin{y_x} < y_x) \wedge%
                             (y_x < 0 \rightarrow \fsin{y_x} > y_x)\\
\textit{$\pi$ Tangent: }   & (y_x < \abst{\pi} \rightarrow \fsin{y_x} < -y_x + \abst{\pi}) \wedge\\%
                           & (y_x > -\abst{\pi} \rightarrow \fsin{y_x} > -y_x - \abst{\pi})\\
\textit{Significant Values: } & (\fsin{y_x} = 0 \leftrightarrow (y_x=0 \vee y_x=\abst{\pi} \vee y_x=-\abst{\pi})) \wedge\\
                           & (\fsin{y_x} = 1 \leftrightarrow y_x=\frac{\abst{\pi}}{2}) \wedge%
                             (\fsin{y_x} = -1 \leftrightarrow y_x=-\frac{\abst{\pi}}{2}) \wedge \\
                           & (\fsin{y_x} = \frac{1}{2} \leftrightarrow (y_x=\frac{\abst{\pi}}{6} \vee y_x=\frac{5*\abst{\pi}}{6})) \wedge \\
                           & (\fsin{y_x} = -\frac{1}{2} \leftrightarrow (y_x=-\frac{\abst{\pi}}{6} \vee y_x=-\frac{5*\abst{\pi}}{6}))
\end{align*}

\subsubsection{Optimization}\ \newline
We use infinite-precision to represent rational numbers.
In our (model-driven) approach,
we may have to deal with numbers with very large numerators and/or denominators.
It may happen that we get such rational numbers from the bad model $\abst{\mu}$ for the variables
appearing as arguments of transcendental functions.
As a result of the piecewise-linear refinement,
we will feed to the SMT(\UFLRA) solver
numbers that have even (exponentially) larger numerators and/or denominators
(due to the fact that \polyApprox uses power series).
This might significantly slow-down the performance of the solver.
We address this issue by
approximating ``bad'' values $\abst{\mu}[x]$ with too large numerators and/or denominators
by using continued fractions~\cite{Nemhauser:1988:ICO:42805}.
The precision of the rational approximation is increased periodically over the number of iterations.
Thus we delay the use numbers with larger numerator and/or denominator,
and eventually find those numbers if they are really needed.
\section{Related work}
\label{sec:related-work}

The approach proposed in this paper is an extension of the approach
adopted in~\cite{tacas17} for checking the invariants of transition
systems over the theory of \emph{polynomial} Nonlinear Real
Arithmetic. %
In this paper we extend the approach to transcendental functions,
with the critical issue of irrational valuations.
Furthermore, we propose a way to prove
SAT without being forced to construct the model.

In the following,
we compare with 
related approaches found in the literature.

\subsubsection{Interval propagation and \deltasat.}
The first approach to SMT(\NTA) was pioneered by
\isat~\cite{DBLP:journals/jsat/FranzleHTRS07}, that carries out
interval propagation for nonlinear and transcendental functions. \isat
is both an SMT solver and bounded model checker for transition
systems.
A subsequent but very closely related approach is the \dreal solver,
proposed in~\cite{gao2012delta}.
\dreal
relies on the notion of
delta-satisfiability~\cite{gao2012delta}, which basically guarantees
that there exists a variant (within a user-specified $\delta$ ``radius'') of the
original problem such that it is satisfiable.
The approach cannot guarantee that the original problem is satisfiable,
since it relies on numerical approximation techniques that only compute safe overapproximations of the solution space.

There are a few key insights that differentiate our approach. First,
it is based on linearization, it relies on solvers for SMT(\UFLRA),
and it proceeds by incrementally axiomatizing transcendental
functions.
Compared to interval propagation, we avoid numerical approximation
(even if within the bounds from \deltasat). In a sense, the precision
of the approximation is selectively detected at run time, while in
\isat and \dreal this is a user defined threshold that is uniformly
adopted in the computations.
Second, our method relies on piecewise linear approximations, which
can provide substantial advantages when approximating a slope --
intuitively, interval propagation ends up computing a
piecewise-constant approximation.
Third, a distinguishing feature of our approach is the ability to
(sometimes) prove the existence of a solution even if the actual
values are irrationals, by reduction to an SMT-based validity check.

\subsubsection{Deductive Methods.}
The \metitarski~\cite{DBLP:journals/jar/AkbarpourP10} theorem prover
relies on resolution and on a decision procedure for
\NRA %
to prove %
quantified
inequalities involving transcendental functions.
It works by replacing transcendental functions with
upper- or lower-bound functions specified by means of axioms
(corresponding to either truncated Taylor series or rational functions derived from continued fraction approximations),
and then using an external decision procedure for \NRA for solving the resulting formulas.
Differently from our approach, \metitarski
cannot prove the existence nor compute a satisfying assignment, while
we are able to (sometimes) prove the existence of a solution even if
the actual values are irrationals.
Finally, we note that \metitarski may require the user to
manually write axioms if the ones automatically selected from a predefined
library are not enough. Our approach is much simpler, and it is
completely automatic.

The approach presented in~\cite{DBLP:conf/frocos/EggersKKSTW11}, where
the \NTA theory is referred to as NLA, is similar in spirit to
\metitarski in that it combines the SPASS theorem
prover~\cite{DBLP:conf/cade/WeidenbachDFKSW09} with the \isat SMT
solver. The approach relies on the SUP(NLA) calculus that combines
superposition-based first-order logic reasoning with SMT(\NTA).
Similarly to our work,
the authors also use a \UFLRA approximation of the original problem.
This is however done only as a first check before calling \isat.
In contrast, we rely on solvers for
SMT(\UFLRA), and we proceed by incrementally axiomatizing
transcendental functions instead of calling directly an \NTA solver.
Another similarity with our work is the possibility of finding solutions in some cases.
This is done by post-processing
an inconclusive \isat answer, trying to compute a certificate for a
(point) solution for the narrow intervals returned by the solver,
using an iterative analysis of the formula and of the computed intervals.
Although similar in spirit,
our technique for detecting satisfiable instances is completely different,
being based on a logical encoding of the existence of a solution as an SMT(\UFLRA) problem.

\newcommand{\Gappa}{\textsc{Gappa}\xspace}
\newcommand{\Coq}{\textsc{Coq}\xspace}
\newcommand{\CoqInterval}{\textsc{Coq.Interval}\xspace}
\newcommand{\NLCertify}{\textsc{NLCertify}\xspace}
\newcommand{\HolLight}{\textsc{Hol-Light}\xspace}

\subsubsection{Combination of interval propagation and theorem proving.}

\Gappa~\cite{gappa, martin2016proving} is a standalone tool and a
tactic for the \Coq proof assistant, that can be used to prove
properties about numeric programs (C-like) dealing with floating-point
or fixed-point arithmetic. Another related \Coq tactic is 
\CoqInterval~\cite{coqinterval}. 
Both \Gappa and \CoqInterval combine interval propagation and Taylor approximations for handling transcendental functions. 
A similar approach is followed also in~\cite{solovyev2013formal}, 
where a tool written in \HolLight to handle conjunctions 
of non-linear equalities with transcendental functions is presented. 
The work uses Taylor polynomials up to degree two. 
\NLCertify~\cite{nlcertify} is another related tool 
which uses interval propagation for handling transcendental functions.
It approximates polynomials with sums of squares 
and transcendental functions with lower and upper bounds using some quadratic polynomials~\cite{allamigeon2013certification}. 
Internally, all these tools/tactics rely on multi-precision
floating point libraries for computing the interval bounds.

A similarity between these approaches and our approach is the use of
the Taylor polynomials. 
However, one distinguishing feature is that we
use them to find lower and upper linear constraints by computing
tangent and secant lines. 
Moreover, 
we do not rely on any floating point arithmetic library, and %
unlike the mentioned approaches, we can also prove the existence of a solution.
On the other hand, some of the above tools employ more sophisticated/specialised approximations
for transcendental functions, 
which might allow them to succeed in proving unsatisfiability 
of formulas for which our technique is not sufficiently precise.

Finally, since we are in the context of SMT, our approach also has the benefits
of being:
\begin{inparaenum}[(i)]
\item fully automatic, unlike some of the above which are meant to be
  used within interactive theorem provers;
\item able to deal with formulas with an arbitrary Boolean structure,
  and not just conjunctions of inequalities; and
\item capable of handling combinations of theories (including
  uninterpreted functions, bit-vectors, arrays), which are beyond what
  the above, more specialised tools, can handle.
\end{inparaenum}

\section{Experimental Analysis}
\label{sec:experiments}

\subsubsection{Implementation.}

The approach has been implemented on top of the \mathsat SMT
solver~\cite{DBLP:conf/tacas/CimattiGSS13}, using the \pysmt library~\cite{pysmt}.
We use the GMP infinite-precision arithmetic library to deal with
rational numbers.
Our implementation and benchmarks %
are available at \url{https://es.fbk.eu/people/irfan/papers/cade17-smt-nta.tar.gz}.

\subsubsection{Setup.}
We have run our experiments on a cluster equipped with 2.6GHz Intel
Xeon X5650 machines, using a time limit of 1000 seconds and a memory
limit of 6 Gb.

We have run \mathsat in two configurations: with and without universal
check for proving SAT (resp. called \mathsat and \mathsatnosat).

The other systems used in the experimental evaluation are
\dreal~\cite{gao2013dreal}, \isatthree~\cite{scheibler2013recent}, and
\metitarski~\cite{DBLP:journals/jar/AkbarpourP10}, in their default configurations (unless
otherwise specified).
Both \isat and \dreal were also run with higher precision than
the default one.
The difference between the two configurations is rather modest and,
when run with higher precision, they decrease the number of \maybesat
answers.
\metitarski can prove the validity of quantified formulae, answering
either valid or unknown. As such, it is unfair to run it on
satisfiable benchmarks.
In general, we interpret the results of the comparison taking into
account the features of the tools.

\subsubsection{Benchmarks.}
We consider three classes of benchmarks.
First, the \emph{bounded model checking (BMC)} benchmarks are the results of
unrolling transition systems with nonlinear and transcendental
transition relations, obtained from the discretization of hybrid
automata.
We took benchmarks from the distributions of \isat, from the discretization
(by way of \hycomp~\cite{hycomp} and \nuxmv~\cite{nuxmv}) of
benchmarks from~\cite{cimatti2016electrical} and from the hybrid model
checkers \textsc{HyST}~\cite{hyst} and \textsc{Hare}\cite{hare}.
Second, the \emph{Mathematical} benchmarks are taken from the \metitarski
distribution. These are benchmarks containing quantified formulae over
transcendental functions, and are all valid, most of them
corresponding to %
known mathematical theorems.
We selected the \metitarski benchmarks without quantifier alternation
and we translated them into quantifier-free SMT(\NTA) problems.
The third class of benchmarks consists of 944 instances from the \dreal distribution
that contain transcendental functions.

Both the mathematical and the \dreal benchmarks
contain several transcendental functions ($\log$, $\cos$, ...)
that are not supported natively by our prototype.
We have therefore applied a preprocessing step that rewrites those functions in terms of $\exp$ and $\sin$.%
\footnote{Sometimes we used a relational encoding: e.g. if $\varphi$ contains $\arcsin(x)$, we rewrite it as $\mksubst{\varphi}{\arcsin(x)}{as_x} \land \sin(as_x) = x \land -\frac{\abst{\pi}}{2} \leq as_x \leq \frac{\abst{\pi}}{2}$, where $as_x$ is a fresh variable.}
\isatthree requires bounds on the variables and it is unable to
deal with the benchmarks above (that either do not specify any bound or specify too wide bounds for the used variables).
Thus, we scaled down the benchmarks so that the variables are constrained in the $[-300,300]$ interval
since for higher bounds \isat raises an exception due to reaching the
machine precision limit.
Finally, for the BMC benchmarks, we run \isatthree in BMC mode,
in order to ensure that its optimized unrolling is activated.

\subsubsection{BMC and Mathematical Results.}

In Table~\ref{table-benchmarks-original}, we present the results.
The benchmarks are classified as either \sat or \unsat when at least one
of the solvers has been able to return a definite answer.
If only \maybesat answers are returned, then the benchmark
is classified as \unknown.
For each tool, we report the number of answers produced within the used resource limits.
For the \maybesat benchmarks,
the numbers in parentheses indicate the instances which have been classified as \sat/\unsat by at least one other tool.
For example, an entry ``87 (32/7)'' means that the tool returned \maybesat for 87 instances,
of which 32 were classified as \sat and 7 \unsat by some other tool.%
\footnote{There was no case in which two tools reported \sat and \unsat for the same benchmark.}

\begin{table}[t!]
\centering
\begin{small}
\begin{tabular}{|l|ccc||ccc|}
\hline
\textbf{Benchmarks} & \multicolumn{3}{c||}{\textbf{Bounded Model Checking (887)}}  & \multicolumn{3}{c|}{\textbf{Mathematical (681)}} \\\hline
\textbf{Result}     &  \textbf{SAT}      &  \textbf{UNSAT}    & \textbf{MaybeSAT}   & \textbf{SAT}          & \textbf{UNSAT}        & \textbf{MaybeSAT}      \\\hline
                                                                                              \hline
\metitarski   &   \na       &    \na       &      \na    &  \na     & \textbf{530} &    \na     \\\hline
\mathsat      & \textbf{72} &    553       &      \na    &  0       &   210        &    \na     \\\hline
\mathsatnosat &   44        & \textbf{554} &      \na    &  0       &   221        &    \na     \\\hline
\isat         &   \na       &    \na       &      \na    &  \na     &   \na        &    \na      \\\hline
\dreal        &   \na       &    392       & 281 (67/23) &  \na     &   285        & 316 (0/253) \\\hline\hline

\textbf{Benchmarks} & \multicolumn{3}{c||}{\textbf{Scaled Bounded Model Checking (887)}} &
                      \multicolumn{3}{c|}{\textbf{Scaled Mathematical (681)}} \\\hline
\textbf{Result}     &  \textbf{SAT}      &  \textbf{UNSAT}    & \textbf{MaybeSAT}   & \textbf{SAT}          & \textbf{UNSAT}        & \textbf{MaybeSAT}      \\\hline
                                                   \hline

\mathsat      & \textbf{84} & \textbf{556} &    \na      & 0       & 215          & \na         \\\hline
\mathsatnosat &   48        & \textbf{556} &   \na       & 0       & 229          & \na         \\\hline
\isat         &   35        &    470       &  87 (32/7)  & 0       & 212          & 137 (0/115) \\\hline
\dreal        &   \na       &    403       & 251 (77/23) & \na     & \textbf{302} & 245 (0/195) \\\hline
\end{tabular}
\end{small}

\caption{Results on the BMC and Metitarski benchmarks.}
\label{table-benchmarks-original}
\end{table}

First, we notice that the universal SAT
technique directly results in 72 benchmarks proved to be
satisfiable by \mathsat, without substantial degrade on the \unsat
benchmarks.
Second, we notice that \metitarski is very strong to deal with its own
mathematical benchmarks,
but is unable to deal with the BMC ones,
which contain features that are beyond what it can handle
(Boolean variables and tens of real variables).%
\footnote{According to the documentation of \metitarski, the tool is ineffective for problems with more than 10 real variables. Our experiments on a subset of the instances confirmed this.}
In the lower part of Table~\ref{table-benchmarks-original}, we present the
results on the scaled-down benchmarks, so that \isatthree can be run.
The results for \dreal and \mathsat are consistent with the ones
obtained on the original benchmarks -- the benchmarks are slightly
simplified for \mathsat, that solves 12 more \sat instances and 2 more \unsat ones,
and for \dreal, that solves 11 more \unsat instances.
The performance of \isat is quite good, halfway between \dreal and
\mathsat on the bounded model checking benchmarks, and slightly lower
than \mathsat on the mathematical ones.
In the BMC benchmarks, \isat is able to solve 35 \sat and 470 \unsat
instances, 102 more than \dreal and 135 less than \mathsat.

The \maybesat results need further analysis. We notice that both \isat
and \dreal often return \maybesat on unsatisfiable benchmarks
(e.g. all the mathematical ones are \unsat).
There are many cases where \dreal returns a \deltasat result, but at
the same time it prints an error message stating that the
numerical precision limit has been reached. Thus, it is unlikely that the
result is actually \deltasat, but it should rather be interpreted as \maybesat
in these cases.\footnote{We contacted the authors of \dreal
  and they reported that this issue is currently under investigation.}

\begin{table}[t!]
\centering
\begin{small}
\begin{tabular}{|l|ccc|}
\hline
\textbf{Benchmarks} & \multicolumn{3}{c|}{\textbf{\dreal (all) (944)}}\\\hline
\textbf{Status}     &  \textbf{SAT}      &  \textbf{UNSAT}    & \textbf{MaybeSAT} \\
\hline\hline
\dreal (orig.)&   \na      & \textbf{102} &    524(3/4) \\\hline
\mathsat      & \textbf{3} &    68        &    \na      \\\hline
\dreal        &   \na      &    44        &     57(3/4) \\\hline
\end{tabular}
\begin{tabular}{|l|ccc|}
\hline
\textbf{Benchmarks} & \multicolumn{3}{c|}{\textbf{\dreal (exp/sin only) (96)}}\\\hline
\textbf{Status}     &  \textbf{SAT}      &  \textbf{UNSAT}    & \textbf{MaybeSAT} \\
\hline\hline
\dreal (orig.)&   \na &    17  &     37 (3/3)   \\\hline
\mathsat      & \textbf{3} &    39       &    \na         \\\hline
\multicolumn{4}{ } \\
\end{tabular}
\end{small}

\caption{Results on the Dreal benchmarks.}
\label{table-benchmarks-dreal}
\end{table}

\subsubsection{\dreal Benchmarks Results.}

The \dreal benchmarks turn out to be very hard. The results are
reported in Table~\ref{table-benchmarks-dreal}, where we show the
performance of \dreal both on the original benchmarks and on the ones
resulting from the removal via pre-processing of the transcendental
functions not directly supported by \mathsat.
The results shows that in the original format \dreal solves many more
instances, and this suggests that dealing with other transcendental
functions in a native manner may lead to substantial improvement in
\mathsat too.
Interestingly, if we
focus on the subset of 96 benchmarks that only contain $\exp$ and $\sin$
(and are dealt by \mathsat without the need of preprocessing),
we see that \mathsat is significantly more effective than \dreal in
proving unsatisfiability, solving more than twice the number of instances
(right part of Table~\ref{table-benchmarks-dreal}).

We conclude by noticing that overall \mathsat solves 906 benchmarks
out of 2512, 127 more than \dreal, the best among the other systems. 
A deeper analysis of the results (not reported here for lack of space)
shows that the performance of the solvers is complementary: the
``virtual-best system'' solves 1353 
benchmarks. This suggests that the integration of interval propagation
may yield further improvements.

\ignore{
\paragraph*{Performance}
\agTODOnote{Where are the cactus plots?}
The cactus plots (reported in the appendix) demonstrate that \mathsat
is slower on the easy (less than one second) benchmarks, due to a
substantial start-up of the Python runtime, but becomes significantly
faster as the benchmarks become harder, ultimately outperforming the
other solvers.\todo[inline]{still to be completed} The scatter plots
demonstrate that our method dominates the other methods.

\todo[inline]{MAKE SURE WE ARE CONSISTENT WITH THE FOLLOWING STATEMENT FROM INTRO:
The results show that our solver is able to solve more benchmarks than
the others, and demonstrates better scalability on the harder
benchmarks.}
}
\section{Conclusion}
\label{sec:conclusion}

We present a novel approach to Satisfiability Modulo the theory of
transcendental functions. The approach is based on an
abstraction-refinement loop, where transcendental functions are
represented as uninterpreted ones in the abstract space SMT(\UFLRA), and
are incrementally axiomatized by means of piecewise-linear functions.
We experimentally evaluated the approach on a large and heterogeneous
benchmark set: the results demonstrates the potential of our approach,
showing that it compares favorably with both delta-satisfiabily and
interval propagation and with methods based on theorem proving.

In the future we plan to exploit the solver for the verification of
infinite-state transition systems and hybrid automata with nonlinear
dynamics, and for the analysis of resource consumption in temporal
planning.
Finally we would like to define a unifying framework to compare
linearization and interval propagation, and to exploit the potential
synergies.

\bibliographystyle{splncs03}
\bibliography{ref}

\end{document}